\documentclass[twocolumn,prl,aps,groupedaddress,nopacs,footinbib,preprintnumbers,floats,floatfix]{revtex4}
\usepackage{graphicx}
\usepackage{verbatim}
\usepackage{mathrsfs}
\usepackage{amssymb}
\usepackage{amsmath}
\usepackage{bm}
\newcommand{\bra}[1]{\left(#1\right)}
\newcommand{\bras}[1]{\left[#1\right]}

\newcommand{\ber}{\begin{eqnarray}}
\newcommand{\eer}{\end{eqnarray}}

\newcommand{\be}{\begin{equation}}
\newcommand{\ee}{\end{equation}}
\newcommand{\ba}{\begin{eqnarray}}
\newcommand{\ea}{\end{eqnarray}}

\newcommand{\ADEPT}{{\sc Adept}}
\newcommand{\BOSS}{{\sc Boss}}
\newcommand{\map}{{\sc Wmap}}
\newcommand{\planck}{{\sc Planck}}

\def\gtorder{\mathrel{\raise.3ex\hbox{$>$}\mkern-14mu
             \lower0.6ex\hbox{$\sim$}}}

\begin{document}

\title{Fundamental Uncertainty in the BAO Scale from Isocurvature Modes}
\author{C. Zunckel$^{1,2}$, P. Okouma$^{3,4}$, S. Muya Kasanda$^{1,4}$, K. Moodley$^{1,4}$ $\&$ B. A. Bassett$^{3,4,5,6}$  \\
\it $^1$ Astrophysics and Cosmology Research Unit, School of Mathematical Sciences, University of Kwazulu-Natal, Durban, 4041, SA\\
\it $^2$ Astrophysics Department, Princeton University, Peyton Hall, 4 Ivy Lane, NJ, 08544, USA\\
\it $^3$ Department of Maths and Applied Maths, University of Cape Town, Rondebosch 7701, Cape Town, SA\\
\it $^4$ Centre for High Performance Computing, CSIR Campus, 15 Lower Hope St., Rosebank, Cape Town, SA\\
\it $^5$ South African Astronomical Observatory, Observatory, Cape Town, SA\\
\it $^6$ African Institute for Mathematical Sciences, Muizenberg, Cape Town, SA}

\begin{abstract}
Small fractions of isocurvature perturbations correlated with the dominant adiabatic mode are shown to be a significant primordial systematic for future Baryon Acoustic Oscillation (BAO) surveys, distorting the standard ruler distance by broadening and shifting the peak in the galaxy correlation function. Untreated this systematic leads to biases that can exceed $10\sigma$ in the dark energy parameters even for {\planck}-level isocurvature constraints.  Accounting for the isocurvature modes corrects for this bias but degrades the dark energy figure of merit by at least 50\%. The BAO data in turn provides extremely powerful new constraints on the nature of the primordial perturbations. Future large galaxy surveys will thus be powerful probes of the earliest phase of the universe in addition to helping pin down the nature of dark energy.
\end{abstract}

\maketitle

\paragraph{Introduction}

Large galaxy surveys fall squarely in the realm of astrophysics and traditionally we think of them as living almost independently of the physics of the early universe. In this {\em Letter} we show that this assumption breaks down rather dramatically in the case of Baryon Acoustic Oscillations (BAO) (for a recent review see \cite{BAOReviews} ). BAO surveys are a key component of the global plan for the next two decades in cosmology because they are believed to provide a robust and powerful statistical standard ruler that can probe dark energy.  They have shown to be robust to a variety of potential systematic effects which only become important at the $\sim 1\%$ level \cite{SE03,SE07}. However, here we show that there is a much more significant ``systematic" arising from the possibility of isocurvature modes correlated with the dominant adiabatic perturbation which may have been generated during the early universe.

To understand this systematic, consider the standard ruler provided by the distance that sound waves can propagate in the primordial plasma. The standard picture based on adiabatic perturbations suggests that the key scale is the sound horizon:
\be
r_{s} = \int^{t_{cmb}}_0 c_s(1+z) dt = \int^{\infty}_{z_{cmb}} \frac{c_s(z')}{H(z')} dz',\nonumber
\label{soundhorizon}
\ee where $c_s(z) = 1/\sqrt{3\bra{1+R_b(1+z)^{-1}}}$ and $R_b = 31500 \, \omega_b\bra{T_{cmb}/2.7 \,{\mbox K}}^{-4}$. The measurement of the late-time clustering of galaxies in the transverse direction probes the angular diameter distance given by $d_A(z) = {r_\perp}(1+z)^{-1}/{\Delta \Theta}$
where $r_{\perp}$ is the intrinsic size of $r_s$ in the transverse direction and $\Delta \Theta$ is the position of the peak in the angular correlation function, while the clustering on a scale $r_{||}$ along the line of sight  probes the Hubble parameter, $H(z) = { \Delta z }/{r_{||}}$.  Measurements of  the angular diameter distance and Hubble parameter in a series of redshift bins using the BAO technique provide an effective probe of the properties of dark energy \cite{BAO_DE_Constraints, SE03}, with prospective constraints on the equation of state of dark energy, $w_0,$ and its evolution $w_a,$ as low as 0.01 and 0.05, respectively, for a future space-based spectroscopic mission (e.g., {\ADEPT} \cite{ADEPT}), with forecasts for current experiments (e.g.,  {\BOSS} \cite{BOSS_white_paper}) at the level of 0.03 and 0.07 respectively.

Systematic effects that affect the position and shape of the Baryon Acoustic Peak (BAP), such as nonlinearity and redshift-space distortions, have been studied and can be treated without a significant impact on dark energy constraints \cite{SE03,SE07}. Here we concentrate on the possibility that the initial conditions were not purely adiabatic.
Isocurvature modes deform this characteristic scale which manifests in the anisotropies in the cosmic microwave background and the clustering of matter. We investigate whether a small admixture of isocurvature modes could potentially degrade constraints on dark energy parameters if allowed for, or bias the measured values if not taken into account. Put more generally, we investigate whether it is possible to decouple the physics of the generation of the primordial density perturbation from the constraints arising from our observations at late times, even with strong prior constraints on the isocurvature modes from CMB data. Constraints from {\map} 3 year data indicate that a 50\% admixture of three isocurvature modes with the adiabatic mode is permitted \cite{Bean06}, whereas forecasts for the {\planck} experiment indicate that isocurvature mode admixtures will be constrained to below the 10\% level \cite{Bucher01}.

What are the possible origins of isocurvature modes? The simplest possibility perhaps is multiple field inflation \cite{early1,early2}, with the curvaton mechanism as a special case \cite{curv}. The resulting isocurvature perturbation is a leading candidate to explain any primordial non-Gaussianity and can, in certain cases, explain the observed asymmetry in the CMB  \cite{Erickcek09}. While the simplest, adiabatic models of inflation are currently preferred \cite{latest2}, it is possible that some isocurvature contamination will be uncovered in future experiments and indeed this would be very fortuitous since it would provide new handles on the physics of the very early universe.  In this letter we argue that allowing for the possibility of isocurvature modes is crucial in future BAO surveys and that as a reward, such surveys can provide a powerful lens on the early universe.

\paragraph{The BAO peak with adiabatic and isocurvature initial conditions}

\begin{figure}[ht!]
\centering
\includegraphics[scale=.6]{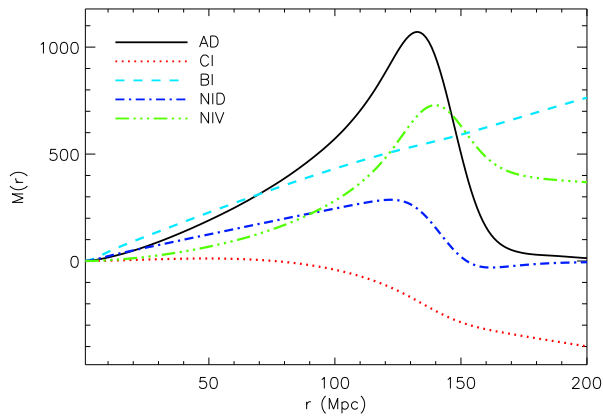}
\caption{Baryon mass profile for adiabatic and isocurvature modes at decoupling, with the correct relative amplitude between the AD (black solid), NID (blue short dot-dashed line)  NIV (green long dot-dashed line), BI (cyan dashed line) and CI (red dotted line) modes.} \label{massprofile_at_dec}
\end{figure}

\begin{figure}[ht!]
\centering
\includegraphics[scale=.6]{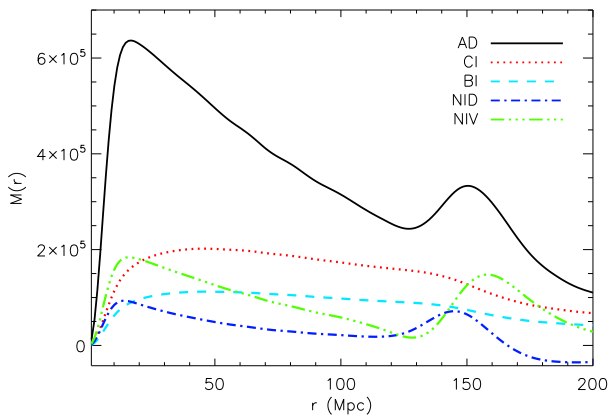}
\caption{Baryon mass profile for adiabatic and isocurvature modes at $z=0$, with the same labelling as in figure 1. For clarity the NID, NIV and CI mass profiles have been multiplied by a factor of two, while the BI mass profile has been multiplied by a factor of six.}
\label{massprofile_today}
\end{figure}

The features of the BAO peak are not only sensitive to the background dynamics, but also to the evolution of the cosmic perturbations, in particular the manner in which the initial relative perturbations between the different species were established. In addition to adiabatic fluctuations, in which the equation of state of the universe is spatially constant and the curvature is perturbed, there is the possibility of regular isocurvature perturbations, in which the equation of state between the different species varies to keep the curvature constant, and of correlations between the adiabatic and isocurvature perturbations \cite{Bucher00}.

Acoustic oscillations in the photon-baryon fluid are described by the photon density evolution equation in harmonic space
\begin{equation}
\label{diff1}
\ddot{\delta}_{\gamma} +\frac{\dot{R}}{1+R}\dot{\delta}_{\gamma}+k^2c_s^2\delta_{\gamma}=  -\frac{2}{3}\left[\frac{\dot{
R}}{1+R}\dot{h}+\ddot{h}\right] \equiv F(k,\tau) , \nonumber
\end{equation}
where $R=\dfrac{R_b}{(1+z)}$ is the baryon-to-photon density ratio, $h$ is the metric field in synchronous gauge, and the dot refers to the conformal time derivative. Solving this equation prior to decoupling we find that
\begin{align}
&\delta_{\gamma}(k,\tau) =A_S \, \sin{kr_s(\tau)} + A_C \cos{kr_s(\tau)} \nonumber\\
\label{deltg_bi}
+ &A_I\int_0^{\tau} (1+R(\tau'))^{1/2}\sin{[kr_s(\tau)-kr_s(\tau')]}F(k,\tau')d\tau',\nonumber
\end{align}
with the initial conditions that define the regular modes given by
\begin{center}
\begin{tabular}{|c|c|c|c|c|c|}
\hline
 & ADIA  & BI & CI & NID & NIV \\
\hline
$A_S$  &  0  & $-\frac{8}{\sqrt{3}k}\Omega_{c,0}$  &
$-\frac{8}{\sqrt{3}k}\Omega_{b,0}$ &  0  & $\frac{4}{\sqrt{3}} \frac{R_{\nu}}{R_{\gamma}}$\\
$A_C$  &   0 & 0   & 0  &  $- \sqrt{3}\,c_s \frac{R_{\nu}}{R_{\gamma}}$  & 0\\
$A_I$    &  $\frac{\sqrt{3}}{k}$ &  $\frac{\sqrt{3}}{k}$  & $\frac{\sqrt{3}}{k}$  & 0   & 0\\
\hline
\end{tabular}
\end{center}
where  $\Omega_{c,0}$ and $\Omega_{b,0}$ are, respectively, the cold dark matter and the baryon densities today, with $R_\nu$ and $R_\gamma$ the fractional energy densities of neutrinos and photons at early times, respectively. The NIV mode starts with a non-zero perturbation in the photon velocity so that it stimulates the $\sin(kr_s)$ harmonic in $\delta_\gamma$, in contrast to the NID mode which has an initial non-zero perturbation in the photon density and excites the $\cos(kr_s)$ harmonic. We note that the adiabatic mode is sourced purely by the gravitational driving term $F(k,\tau),$ which is constant on large scales and at early times, so that the adiabatic solution excites a $\cos(kr_s)$ harmonic, similar to the NID mode. At later times, though, this approximation breaks down when the mode enters the horizon, and the adiabatic solution deviates from a pure cosine mode, with the consequence that its acoustic peaks are offset in phase from the NID peaks.  In the transition to matter domination, the gravitational driving term switches on as a source for the CI and BI modes, whereas its contribution to the NID and NIV modes remains negligible.

The different harmonics stimulated by the various modes result in a relative shift of the BAO peak position. We observe the same shift in the series of acoustic peaks exhibited in the cosmic microwave background anisotropy spectrum \cite{Dunkley05}, though in this case it is recombination, as opposed to decoupling, that sets the acoustic scale. To quantify the shifts in the BAO peak of the isocurvature modes relative to the adiabatic mode, in figure
\ref{massprofile_at_dec} we plot the baryon mass profile \cite{Eisenstein07} of the various modes at decoupling, defined by
\begin{equation}
M_{b}(r,z)=\int T_{b}(k,z) \frac{\sin{kr}}{kr} k^2 r^2 e^{-k^2\sigma^2/2}dk,\nonumber
\label{eq1}
\end{equation}
where $T_{b}= \delta_b/k^2$ is the baryon transfer function. Prior to decoupling and on large scales, $\delta_b$ is exactly ${3\over 4} \delta_\gamma$ for all modes except the BI and NID modes in which the baryon and photon density perturbations differ by a constant, respectively, $1$ and $R_\nu/R_\gamma$.

The mass profile captures the evolution of an initial point-like perturbation at the origin, though in practice we use a narrow gaussian, with width $\sigma^{-1}$ in $k$ space. The initial density perturbation expands out as a spherical wave at the sound speed \cite{Bashinsky01} so that at decoupling there is an excess of baryons at the sound horizon scale, $r_s \approx c_s \tau_{\rm dec}$. Note that even though the photons decouple from the baryons at $z  \approx 1080,$ it is only at $z \sim 500$ that the baryons stall, due to the fact that the growing mode is dominated by the velocity field on small scales, which does not decay instantaneously after decoupling. This sets the scale for the low-redshift BAO peak, which is shown in figure \ref{massprofile_today}. We observe that the BAO feature for the isocurvature modes differs in shape and position from the adiabatic BAO peak. In the case of the NID and NIV modes there is a pronounced peak which is offset from the adiabatic BAO peak due to the coupling to different harmonics, as described above. In the case of the BI and CI modes the acoustic wave has merely imprinted a ripple onto the homogeneous sea of baryons at decoupling, which evolves into a knee in the baryon mass profile at the BAO scale at late times. It is also interesting to note that the amplitude of the mass perturbation at the origin is much smaller for the isocurvature modes because the curvature, and thus the mass fluctuation, is initially unperturbed. It is clear that small admixtures of the isocurvature modes and their cross-correlations can distort the shape and location of the adiabatic BAO peak.

\paragraph{Dark energy constraints}

\begin{figure}
\begin{center}
\begin{tabular}{c}
 \includegraphics[trim = 10mm 21mm 0mm 0mm, scale=0.7, angle=0]{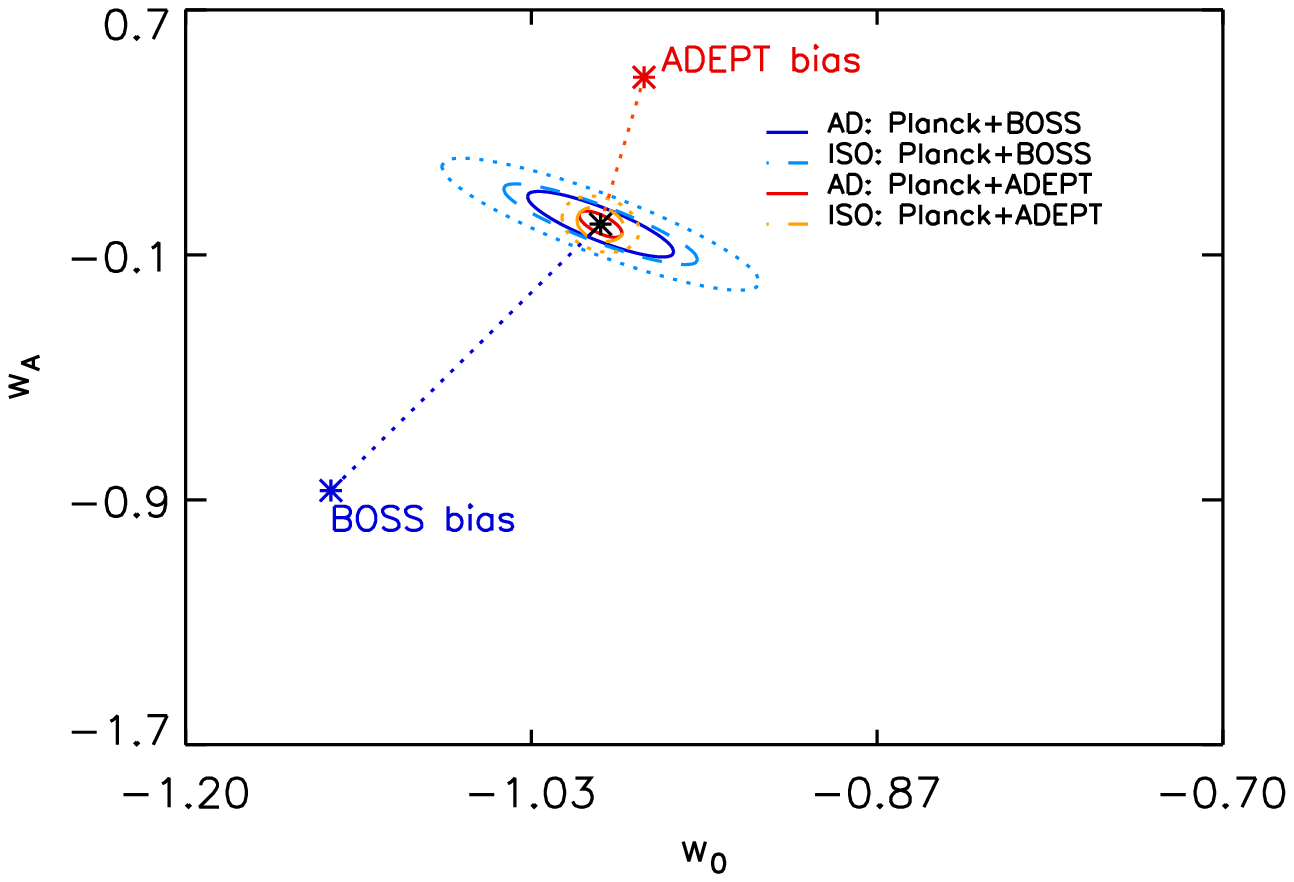} \\
  \includegraphics[trim = 10mm 10mm 93mm 185mm, scale=0.38, angle=0]{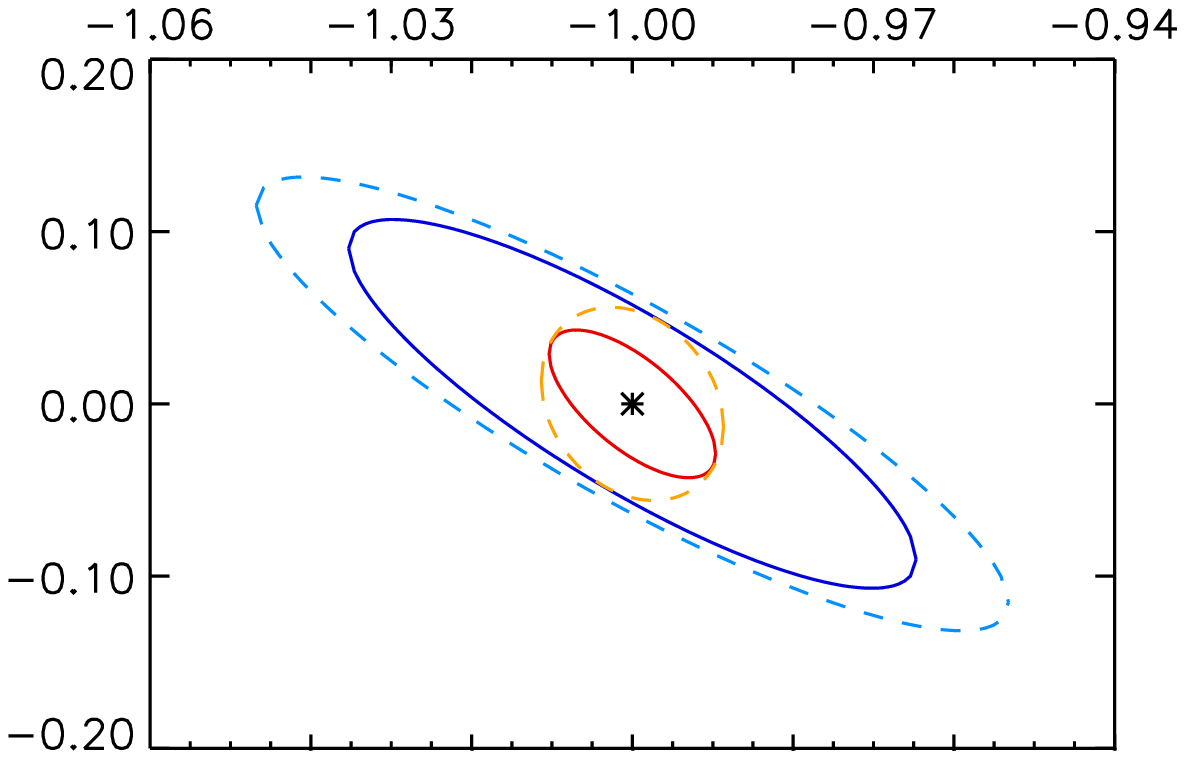} 
\end{tabular}
 \end{center}
\caption{ $1\sigma$ error ellipses for $(w_0, w_a)$ from the Fisher matrix calculation for the {\BOSS} (blue) and {\ADEPT} (red) experiments with the {\planck} data as a prior, 
assuming adiabaticity (solid) and for the fully correlated isocurvature case (dashed). The fiducial model is marked by a black star. The average biases in the dark energy parameters that could potentially incurred as a result of the incorrect assumption of adiaticity are shown for the two experiments by the dotted lines.  All other parameters have been marginalized over. An enlarged version of the error ellipses is shown in the insert. }  
\label{AD_vs_ISO}
\end{figure}

Assuming that the variations in the cosmological parameters (including those modeling isocurvature) are small, we can model the likelihood function of a dataset as a multivariate Gaussian centered on a fiducial adiabatic $\Lambda$CDM Universe.  Based on the noise estimates for the {\BOSS} and {\planck} experiments, we can compute estimates of the errors on the cosmological parameters using a Fisher matrix formalism, by perturbing the cosmology around the fiducial model. When perturbing the dark energy model away from $\Lambda$, we allow for dynamics and parameterize its equation of state using $w(a) = w_0 + (1-a) w_a$ where $a=1/(1+z)$ 
\cite{Chevallier:2000qy,linder-2003-90}. To model deviations away from adiabaticity, we adopt the isocurvature parameterization implemented in \cite{Moodley04}, where the different modes and their cross-correlations are described by 9 parameters, measuring the fractional contributions of the various correlations (auto and cross) to the overall total power spectrum.  Note that for the computation of the {\planck} Fisher matrix, we follow \cite{DEFT} and re-introduce the strict geometric degeneracy between the dark energy density $\Omega_X$ and $w_0$, $w_a$ which may be artificially broken in the standard Fisher matrix computation, leading to under-estimates of errors.

We are concerned with 
how well the equation of state parameters $w_0$ and $w_a$ can be measured. 
In order to quantify the constraining power of the data, we compute the Dark Energy Task Force (DETF) figure of 
merit (FoM), which is defined as the reciprocal of the area in
the $w_0 - w_a$ plane, enclosing the $95\%$ confidence limit (CL) region \cite{DEFT}. 
We are concerned with the change in the FoM when isocurvature is introduced relative to the case of pure adiabaticity.

We compute the potential errors on $w_0$ and $w_a$ for the case of pure adiabaticity and 
for the scenario where all isocurvature modes and their cross-correlations
are admitted, while marginalizing over all other cosmological parameters. The results for both 
the {\BOSS} and {\ADEPT} experiments are shown in the first and last rows of table \ref{table2}. The {\BOSS} FoM is found to increase by $50\%$ when this additional freedom is introduced in the initial 
conditions, while the {\ADEPT} FoM degrades by $90\%$. 
These changes are illustrated in figure 
\ref{AD_vs_ISO} which compares the $68\%$ confidence regions for the adiabatic and the fully 
correlated isocurvature cases.    Evidently, no single mode and its correlation are 
responsible for the change  in the allowable $(w_0,w_a)$ region, but rather a mixture of all extra 
degrees of freedom.   Note that the forecasted errors on $w_0$ and $w_a$ for the adiabatic case in our 
analysis are considerably smaller than those reported in \cite{BOSS_white_paper}, likely due to our assumption of spatial flatness, but what is important is the relative change in the FoM.

\begin{table}
\begin{center}
\begin{tabular}{|ccccc|}
\hline
Experiment &  {\BOSS} & &  {\ADEPT} & \\ 
\hline
Modes &  $w_0$ & $w_a$ & $w_0$ & $w_a$ \\
\hline
AD    &   0.023 & 0.071 &  0.0068&  0.028\\
AD+CI+$\langle AD,CI\rangle$ &   0.024 &  0.071 & 0.0069 & 0.030 \\
AD+NID+$\langle AD,NID\rangle$&  0.024 &  0.072  & 0.0069 & 0.028 \\ 
AD+NIV+$\langle AD,NIV\rangle$ &  0.025 &  0.072  & 0.0069 & 0.031\\
ISO (ALL) &  0.031 &  0.087  & 0.0075 & 0.037\\
\hline 
\end{tabular}
\end{center}
\caption{Table summarizing the constraints on $(w_0,w_a)$ for adiabatic and an admixtures of uncorrelated 
adiabatic and isocurvature modes, marginalizing other all other parameters, for the {\BOSS} and {\ADEPT} experiments.  
The fiducial model assumes adiabaticity.  }
\label{table2}
\end{table}

The question that we wish to ask is what bias in the estimates of the dark energy parameters
could potentially be induced by incorrectly assuming adiabaticity? For a Gaussian-distributed
likelihood function, it can be shown that the linear bias in a set of parameters that we wish to
constrain, $\delta \theta_i,$ due to erroneous values of a set of fixed parameters, $\delta \phi_j,$ is  \cite{Taylor}
\be
\delta \theta_i = -\bras{F^{\theta \theta}}_{ik}^{-1} F^{\theta \phi}_{kj}\delta \phi_j
\ee where $F^{\theta \theta}$ is the Fisher sub-matrix for the parameters we wish to constrain and
$F^{\theta \phi}$ is a Fisher sub-matrix constructed from the product of the derivatives of the
power spectrum with respect to the parameters being constrained and those which are being fixed.
In our case $j$ labels the nine isocurvature mode amplitudes, incorrectly fixed to zero, $k$ labels the eight cosmological parameters that are biased, and $i$ labels the subset of two dark energy parameters whose bias is of interest to us.
In order to set $\delta \phi_j$, we diagonalize the combined {\planck} and large-scale structure (LSS) Fisher matrix and select the eigenvector, ${\bf e}_i$ with the
smallest eigenvalue $\lambda_i$. This corresponds to the direction in parameter space which is least constrained by the data.
We then take $\delta \phi_j = \sqrt{\frac{19.2}{\lambda_i}}{\bf e}_i$. 
For the {\BOSS} experiment, we find that $\delta w_0 = -0.044$ and $\delta w_a = 0.12$, while for {\ADEPT} the biases are found to be $\delta w_0 = -0.0095$ and $\delta w_a = -0.061$. 

However this is only one particular direction
that weakly constrains all the parameters, not necessarily the dark energy parameters. 
In order to explore the full range of the bias,
we use a set of 10,000 random linear combinations of the eigenvectors to compute $\delta \phi_j$ and the corresponding biases.
We find that the mean biases are $\mu(\delta w_0)= -0.13$ and  $\mu(\delta w_a) = -0.87$ with $\sigma(\delta w_0) = 0.072$ and 
$\sigma(\delta w_a) = 0.39$. 
The implication is that if the initial conditions of our Universe are comprised of a 
sub-dominant contribution from isocurvature modes (within the $1\sigma$ constraints from {\planck} and the selected LSS survey), 
the assumption of adiabaticity could lead to an incorrect $6\sigma$ detection of non-$\Lambda$ 
dark energy model or a $12\sigma$ false claim of dynamics.  Alternatively, $\Lambda$ could be found to be consistent with the data when in 
fact $w(z)\neq -1$.  The potential bias incurred by the adiabatic assumption in the case of a more advanced BAO experiment 
such as {\ADEPT} is somewhat smaller for $w_0$, with $\mu(\delta w_0)= 0.021$ and $\sigma(\delta w_0) = 0.0072$, while the measurement of $w_a$ could be inaccurate at the level of $17\sigma$ with $\mu(\delta w_a) = 0.48$  and  $\sigma(\delta w_a) = 0.18$.
The reduction in the bias is encouraging as one would expect that with an increase in the constraining power of the survey comes a 
higher risk of making false claims. We note that the estimates in this paper so far are optimistic because 
we assume perfect knowledge of the galaxy bias and redshift distortions. 

Although allowing for isocurvature modes degrades the dark energy constraints relative to the perfectly adiabatic case, there is a powerful positive. The volume of the 9-dimensional isocurvature Fisher ellipse is roughly $2-4 \times 10^9$ smaller than that from {\planck} alone, showing that using CMB plus BAO data in union provides exceptionally good constraints on the early universe relative to the CMB alone. More specifically, we find that the error bars on the isocurvature parameters decrease by 30$\%$ to as much as 100$\%$ for certain modes when the LSS data (either {\BOSS} or {\ADEPT}) is added to the {\planck} data.  

\paragraph{Conclusions}
With forecasted constraints on dark energy from BAO experiments at the level of a few percent made possible by the large volumes probed by the most recent generation of redshift surveys, it is important to explore the full spectrum of possible BAO systematics.  In this \emph{Letter}, we have revisited the assumption of adiabatic initial conditions.  We have found that the admission of isocurvature admixtures alters the position of the BAP, and by assuming adiabaticity we run the risk of incorrectly attributing a shift in the peak (away from the predicted value in a $\Lambda$CDM model) to the presence of dark energy. 

Ignoring isocurvature modes can substantially bias the estimates of the dark energy parameters, leading to a $6\sigma$ ($3\sigma$)  incorrect measurement of $w_0$ or as much as a $12\sigma$ ($17\sigma$) bias in $w_a$ for {\BOSS} ({\ADEPT}) . Allowing for general initial conditions removes this bias at the expense of an increase in the $2\sigma$ region in $(w_0,w_a)$ space by $50\%$ ($90\%$) for the {\BOSS} ({\ADEPT}) survey, indicating that the assumption of adiabaticity can lead to an under-estimation of the errors on the dark energy parameters. On the other hand, BAO data substantially improves the constraints on isocurvature modes.  

\paragraph{Acknowledgements}
CZ is funded by a NRF/DST (SA) Innovation Fellowship and a National Science Foundation (USA) fellowship under grant PIRE-0507768. PO is funded by the SKA (SA). SMK received support via the Meraka Institute via funding for the South African Centre for High Performance Computing (CHPC). BB and KM acknowledge support from the National Research Foundation, SA.

\end{document}